\documentclass[preprint,showpacs,amsmath,amssymb]{revtex4}

\usepackage{epsfig}

\begin{document}

\title{Achieving Anisotropy in Metamaterials made of Dielectric Cylindrical Rods}

\author{L. Peng$^{1,2}$}

\author{Lixin Ran$^{1}$}
\email{ranlx@zju.edu.cn}
\author{N.~A. Mortensen$^{2}$}
\email{asger@mailaps.org}

\affiliation{$^{1}$Department of Information and Electronic Engineering, Zhejiang University, Hangzhou 310027, P.R. China\\
$^{2}$Department of Photonics Engineering, Technical
University of Denmark, DTU-building 345 west, DK-2800
Kongens Lyngby, Denmark}

\date{\today}

\begin{abstract}
We show that anisotropic negative effective dispersion relation can be achieved in pure dielectric rod-type metamaterials by turning from the symmetry of a square lattice to that of a rectangular one, i.e. by breaking the rotation symmetry of effective homogeneous medium. Theoretical predictions and conclusions are verified by both numerical calculations and computer based simulations. The proposed anisotropic metamaterial, is used to construct a refocusing slab-lens and a subdiffraction hyperlens. The all-dielectric origin makes it more straightforward to address loss and scaling, two major issues of metallic structures, thus facilitating future applications in both the terahertz and optical range.
\end{abstract}


\maketitle

%%%%%%%%%%%%%%%%%%%%%%%%%%%%%%%%%%%%

In recent years, metamaterials (MMs) made of a stack of small structural elements have attracted substantial attention. The inherent ability of high-index structures to support strong subwavelength resonances allows the associated MMs to behave as effective media with various unusual macroscopic electromagnetic (EM) properties, such as negative permittivity/permeability and negative refraction~\cite{Pendry:1996,Pendry:1999,Shelby:2001}. With peculiar collective EM characteristics, MMs can be used to fabricate innovative devices like subwavelength refocusing lenses~\cite{Pendry:2000} and invisibility cloaks~\cite{Pendry:2006}.

Currently, one of the most important issues on MMs is to scale the MM unitcells down in size to allow operating a yet higher frequencies, i.e. terahertz and optical regions~\cite{Liu:2008,Valentine:2008}. In this context, dielectrics rather than metals might be preferred by MMs, since the dispersive metals would be non-ideal and exhibit significant loss, thus compromising MMs design and application at high frequencies. Till now, great efforts have been devoted to developing homogenization theory for the design and fabrication of pure dielectric MMs. Rod-type dielectric resonators with high permittivity have been used to fabricate left-handed (LH) MMs in the microwave range and the terahertz range~\cite{Peng:2007,Schuller:2007}. Recent scaling analysis have also pointed out that LH behaviors at optical frequencies would also be possible by use of rod-type MMs made of silicon~\cite{Vynck:2009}. Rod-type dielectric MMs with an isotropic density of rods (supported e.g. by an ordered or disordered square-lattice of rods) will commonly possess an isotropic dispersion in the two dimensions~\cite{Peng:2007,Vynck:2009}. While isotropic dispersion is of advantageous in a number of cases, there is however also a strong call for realizing anisotropic MMs for practical applications, as is in general required for a transformation medium.

In this paper, we report on both theoretical and numerical studies of the dielectric rod-type MMs with asymmetrical shaped unit cells. General closed-form expressions for the effective constitutive parameters are obtained from a macroscopic viewpoint, i.e. in the long wavelength limit. We show that the dispersion relationship of such dielectric MMs would be naturally spatially anisotropic with negative refraction, supported by parameters extracted from numerical computation and EM phenomena in other simulations. By using this kind of anisotropic rod-type MMs, a farfield slab-lens with different refocusing distance is conveniently realized. Moreover, this kind of MMs can also be implemented to fabricate a slab-type hyperlens with sub-diffraction limit. Both of these designs are supported by full-wave simulations.

\begin{figure}[b!]
\begin{center}
\epsfig{file=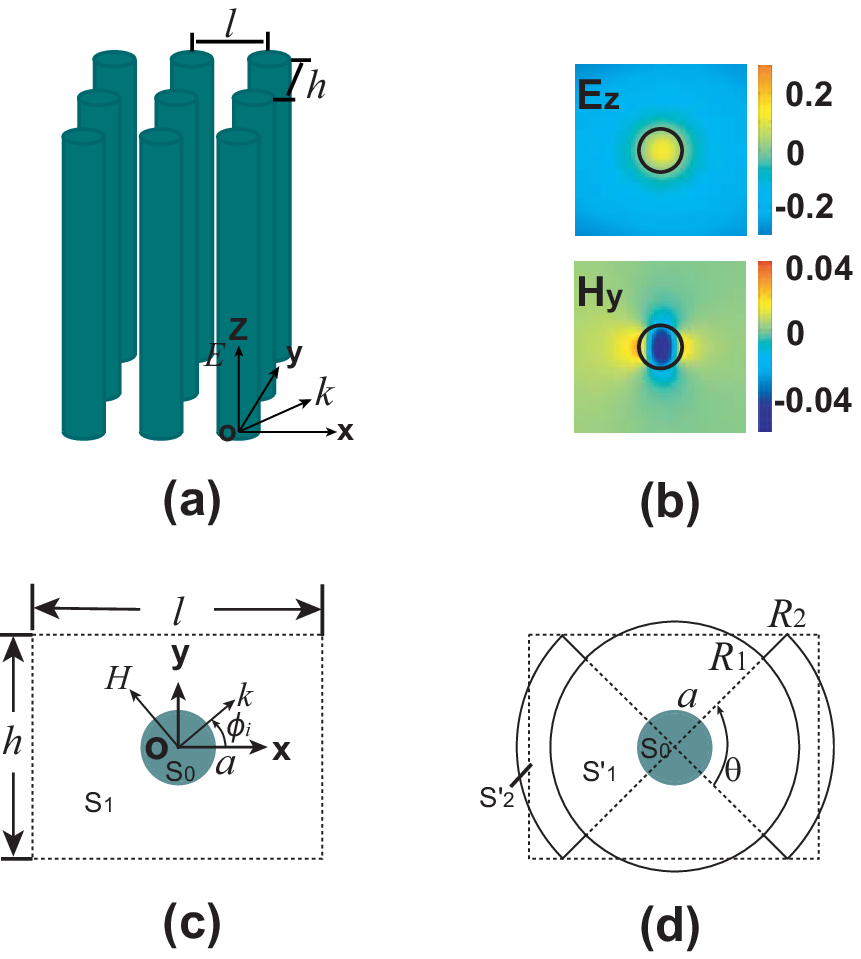, width=0.7\columnwidth,clip}
\end{center}
\caption{(color). (a) The configuration of rod-type dielectric metamaterials. (b) The fundamental electric and magnetic resonant modes in a dielectric cylinder with $z$ polarization incidence; (c) The cross section schematic of rod-type metamaterial unit cell and (d) its simplified integral region.}
\label{fig1}
\end{figure}

We start by considering the array composed of standard cylindrical resonators, as shown in Fig.~\ref{fig1}(a). For simplicity, we emphasize the ordered case, but the composite could be disordered, since the low-frequency macroscopic EM properties of this kind metamaterial do not rely on band structure phenomena, as they are mainly determined by the low-lying Mie resonances of the isolated cylinders with large permittivity~\cite{Peng:2007,Vynck:2009}. As discussed in Ref.~\cite{Peng:2007}, the displacement current distribution inside the cylinder in the basic two Mie resonance modes would be similar to that of the conductive current in metallic rod and Split-ring Resonator (SRR), thus also supporting effective negative permittivity/permeability of the metamaterial. Fig.~\ref{fig1}(b) shows the electric field and magnetic field distributions associated with two first basic Mie resonance modes, respectively. When the dielectric cylinders are arranged, in a regular lattice or in a disordered fashion, the whole composite will exhibit isotropic EM responses provided that the average density of rods is the same for e.g. the $x$ and the $y$ directions. Spatial anisotropy may be introduced by relaxing this isotropic density, thus breaking the effective rotational symmetry of the long-wavelength effective medium.

Here, without loss of generality, rod-type dielectric metamaterial would have a rectangular shaped unit cell, as shown in Fig.~\ref{fig1}(c). Basically, to treat this kind of rod-type composite as an effective medium, the typical dimensions of a single unit cell have to be much smaller than the wavelength in free space~\cite{Peng:2007}. The effective parameters, i.e. ${\boldsymbol\epsilon}_{\rm eff}$ and ${\boldsymbol \mu}_{\rm eff}$, are defined in Ref.~\cite{Peng:2007} as $\big<{\boldsymbol D}\big>={\boldsymbol\epsilon}_{\rm eff}\big<{\boldsymbol E}\big>$ and $\big<{\boldsymbol B}\big>={\boldsymbol \mu}_{\rm eff}\big<{\boldsymbol H}\big>$. From Mie scattering theory, all the correlated fields can formally be expressed by superposition of vector cylindrical waves~\cite{Tsang:2000}. However, for the rods array in two dimensions, when the exciting electric field is polarized along the axis, i.e. the $z$-direction in Fig.~\ref{fig1}(c), the practical geometry of the lattice complicates a closed-form analytical solution for the integration over the rectangular shaped area $S_{1}$. Although numerical evaluations can make precise estimations, we will for the ease of analysis approximate the integral over the rectangular domain by performing the integrals on an equivalent region formed by $S_{0}$, $S'_{1}$ and $S'_{2}$, see Fig.~\ref{fig1}(d). In this way simplify the problem, yet still capturing the correct physics. With these approximations, if $a$ is much smaller than the free wavelength in vacuum, in which case high-order scattering is very weak, straightforward algebraic operation would eventually give
\begin{equation}
\label{eq:epsilon}
\frac{\epsilon_{{\rm eff},z}}{\epsilon_{0}}=\frac{S-\pi a^{2}}{S}\left[1-\frac{2i\epsilon_{p}J_{1}(k_{p}a)}{k_{p}c_{0}}\frac{E_{0}}{X_{z}}\right],
\end{equation}
with $X_{z}\approx \pi a^{(N)}_{0}\left[\rho H^{(1)}_{1}(k\rho)\right]^{R_1}_{a}+\pi E_{0}\left[\rho J_{1}(k\rho)\right]^{R_1}_{a}+\theta a^{(N)}_{0}\left[\rho H^{(1)}_{1}(k\rho)\right]^{R_2}_{R_1}+\theta E_{0}\left[\rho J_{1}(k\rho)\right]^{R_2}_{R_1}$, $c_{n}=\sqrt{\frac{\epsilon_{p}}{\mu_{p}}}H^{(1)}_{n}(ka)J'_{n}(k_{p}a)-H'^{(1)}_{n}(ka)J_{n}(k_{p}a)$ and $a^{(N)}_{n}=\frac{E_{0}}{c_{n}}\left[J'_{n}(ka)J_{n}(k_{p}a)-\sqrt{\frac{\epsilon_{p}}{\mu_{p}}}J_{n}(ka)J'_{n}(k_{p}a)\right]$. $a$, $R_1$, $R_2$, and $\theta$ are parameters depicted in Fig.~\ref{fig1}(d). Similarly, we find
\begin{equation}
\label{eq:mu}
\frac{\mu_{{\rm eff},x(y)}}{\mu_{0}}=\frac{S-\pi a^{2}}{S} \left\{1+\pi a\left[J_{1}(ka)+H^{(1)}_{1}(ka)\frac{a^{(N)}_{1}}{E_{0}}\right]\frac{E_{0}}{Y_{x(y)}}\right\},
\end{equation}
with $Y_{x(y)}=\pi a^{(N)}_{1}\left[\rho H^{(1)}_{1}(k\rho)\right]^{R_1}_{a}+\pi E_{0}\left[\rho J_{1}(k\rho)\right]^{R_1}_{a}+\frac{E_0}{k}(\theta\pm\sin\theta)\left[J_{0}(k\rho)\right]^{R_2}_{R_1}-\frac{E_0}{k}(\theta\mp\sin\theta)\left[J_{0}
(k\rho)\right]^{R_2}_{R_1}+E_0(\theta\mp\sin\theta)\left[\rho J_{1}(k\rho)\right]^{R_2}_{R_1}+\frac{a^{(N)}_{1}}{k}(\theta\pm\sin\theta)\left[H^{(1)}_{0}(k\rho)\right]^{R_2}_{R_1}-\frac{a^{(N)}_{1}}{k}
(\theta\mp\sin\theta)\left[H^{(1)}_{0}(k\rho)\right]^{R_2}_{R_1}+a^{(N)}_{1}(\theta\mp\sin\theta)\left[\rho H^{(1)}_{1}(k\rho)\right]^{R_2}_{R_1}$.

\begin{figure}[t!]
\begin{center}
\epsfig{file=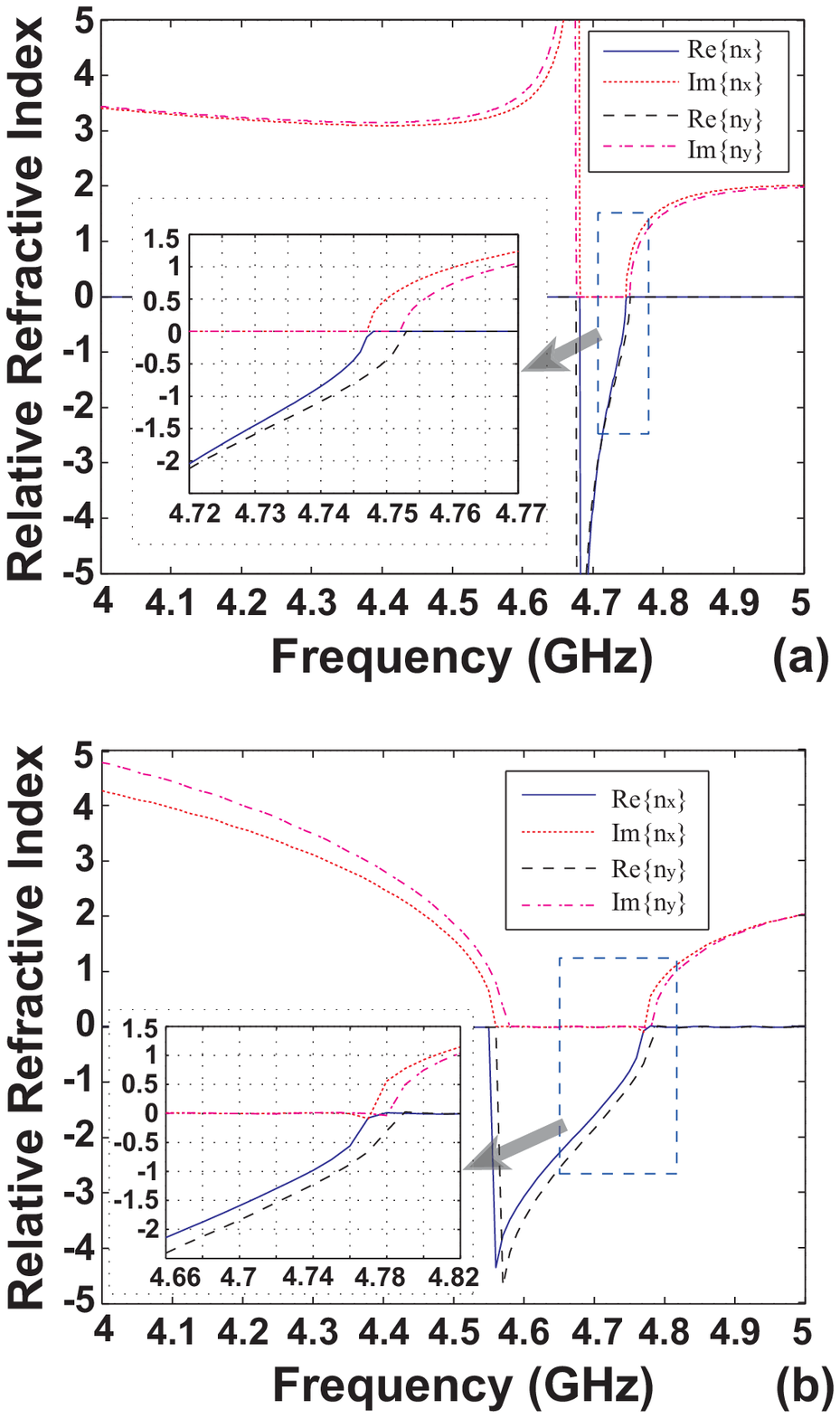, width=0.6\columnwidth,clip}
\end{center}
\caption{Effective refractive index in different directions of anisotropic rod-type metamaterial. (a)Theoretical estimation; (b)Data extracted from numerical simulation. In the calculation and simulation, $a$=1mm, $l$=7.4mm, $h$=7mm, and the relative dielectric constant of the rod is 600.}
\label{fig2}
\end{figure}

As an example, we include numerical evaluations of the above expressions for a practical case. We assume that the dielectric rods are made of Ba$_{0.5}$Sr$_{0.5}$TiO$_{3}$ (BST) ceramic, which has a dielectric constant around 600 in the microwave range~\cite{Peng:2007}. It has previously been shown that a MM made from rods of this ceramic will have a pass band with negative refractive index~\cite{Peng:2007,Vynck:2009}. For the dimensions of the asymmetric lattice we consider  $l$=7.4mm and $h$=7mm, which will support an anisotropic dispersion relation, as compared to isotropic property of the corresponding square lattice. Fig.~\ref{fig2}(a) shows the calculated effective refractive index $n_{\rm eff}=\sqrt{\epsilon_{\rm eff}}\sqrt{\mu_{\rm eff}}$ in both $x$ and $y$ directions. Considering the pre-conditions of our homogenization, we only focus on those frequencies at which the absolute value of the refractive index is small, see the inset of Fig.~\ref{fig2}(a). The difference between $n_x$ and $n_y$ indicates the anisotropic property of this MMs. In addition, we have also performed wave simulations and extract the effective refractive index in the two directions by a retrieval method~\cite{Chen:2004}. The extracted data is shown in Fig.~\ref{fig2}(b). Obviously, Fig.~\ref{fig2}(a) is qualitatively consistent with Fig.~\ref{fig2}(b).

For an effective medium, the dispersion relation of an effective medium could formally be written in an analytical form like $\frac{k^{2}_{x}}{n^{2}_{x}}+\frac{k^{2}_{y}}{n^{2}_{y}}=k^{2}_{0}$, with $n_x$ and $n_y$ being the refractive index in $x$ and $y$ directions, respectively. Though the different effective refractive index associated with different directions is responsible for the anisotropic EM response, it is quite difficult to verify the complete dispersion curve by a retrieval method as proposed by Ref.~\cite{Chen:2004} because of the rotational asymmetry. Here, we instead illustrate the anisotropic property, by the example of a device which require an anisotropic MM, i.e. refocusing of a point source by a slab lens.

\begin{figure}[t!]
\begin{center}
\epsfig{file=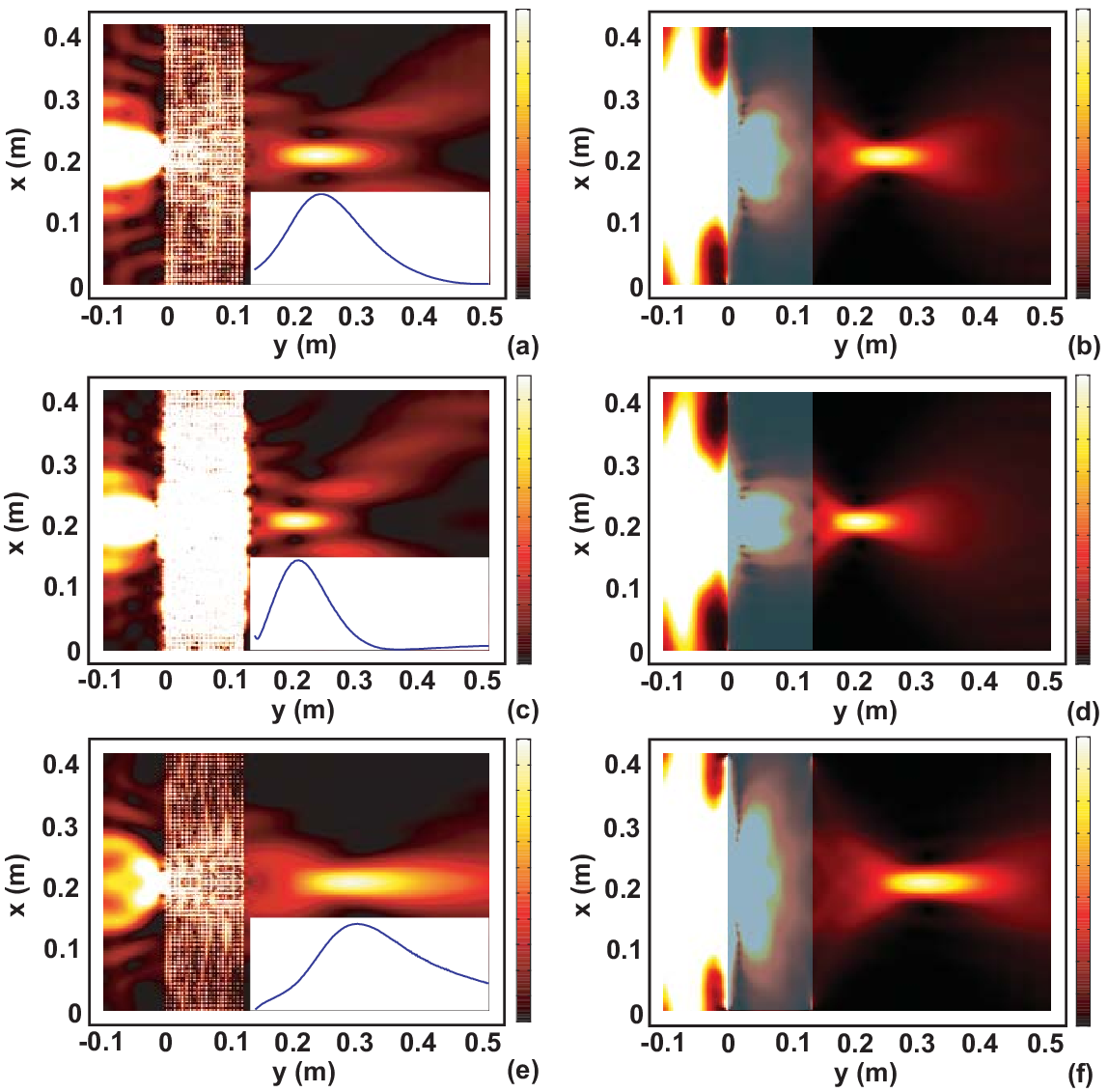, width=1\columnwidth,clip}
\end{center}
\caption{PFDD in the refocusing by three different slab lenses made of pure dielectric metamaterials. The periodicities in $x$ and $y$ directions are (a)7mm$\times$7mm,(c)7mm$\times$7.4mm,(e)7mm$\times$6.4mm; the widths of the three superlens are (a)0.126m,(c)0.1258m,(e)0.128m. In the three simulations, the operation frequencies change slightly, respectively. The insets of (a), (c) and (e) are the cross section of the power flow density distribution at $y$=0.21m. Subfigures (b), (d) and (f) are the PFDD in the refocusing by three ideal anisotropic slablens with negative refractive index. The region ($0\leq y\leq 0.128m$) is occupied by the slablens for (b), (d) and (f).}
\label{fig3}
\end{figure}

As known, a slab lens with refractive index of -1 can make sub-wavelength refocusing in free space~\cite{Pendry:2000}. As an effect of a compensating bilayer, a general slab lens with sub-wavelength resolution can be realized by anisotropic negative refractive index metamaterials (NIMs) whose dispersion relation has a form of $k^{2}_{x}+\frac{k^{2}_{y}}{r^{2}}=k^{2}_{0}$, where $r=(d'-d)/d$ is relating to the distance between source and its image ($d'$) vs. width of the slab lens ($d$)~\cite{Schurig:2005}. By using the rod-type dielectric metamaterial, such a slab lens would be easily designed and constructed. For simplicity, ordered metamaterials are adopted here. The parameters for the first slab lens are $a$=1 mm, $l$=7 mm, and $h$=7 mm, with the rod's relative dielectric constant of 600. Since the dispersion relation of this slab lens would be isotropic, a refocused image can be found at a distance $d'=2d$, see the power flow density distribution (PFDD) in Fig.~\ref{fig3}(a). Next, we keep $l$ unchanged, but vary $h$ in the $y$ direction. In case (\emph{I}), $h$ is increased to 7.4 mm, and from the PFDD in Fig.~\ref{fig3}(c) we see that the source is refocused again at $d'<2d$ with a little down shifting of operation frequency. In case (\emph{II}), $h$ is decreased to 6.4 mm, in which case the refocused image is found at a distance $d'>2d$. For comparison, we have also simulated the PFDD of refocusing by an ideal negative index anisotropic slab lens, see Fig.~\ref{fig3}(b) (isotropic material with $n\approx -1$), and panels~\ref{fig3}(d) and \ref{fig3}(f) is for an anisotropic material with $k^{2}_{x}+1.4^{2}k^{2}_{y}=k^{2}_{0}$ and $k^{2}_{x}+\frac{k^{2}_{y}}{1.5^{2}}=k^{2}_{0}$, respectively.

Clearly, from Fig.~\ref{fig3}, the refocusing of a MM slablens is governed by the intrinsic effective negative constitutive relations. However, it is very hard to obtain subwavelength resolution in such an imaging case, since its effective impedance is generally mismatched with that of the free space~\cite{Peng:2007}, leading to inevitable reflection at the interface. Fortunately, there is an alternative way to achieve sub-wavelength resolution with the current MMs. From Fig.~\ref{fig2}, the effective refractive index in one direction will turn imaginary at some frequencies just beyond the upper bound of the double negative band, i.e. $n_x$ is imaginary at 4.78~GHz in Fig.~\ref{fig2}(b), while $n_y$ stays real. Thus, the MMs' effective dispersion would be hyperbolic in a narrow frequency band, e.g. from 4.77~GHz to 4.79~GHz in Fig.~\ref{fig2}(b). With these characteristics, this MM can be used to form a hyperlens which can distinguish sources with subwavelength separation placed close to the interface~\cite{Salandrino:2006,Jacob:2006,Smolyaninov:2007,Liu:2007}.

\begin{figure}[t!]
\begin{center}
\epsfig{file=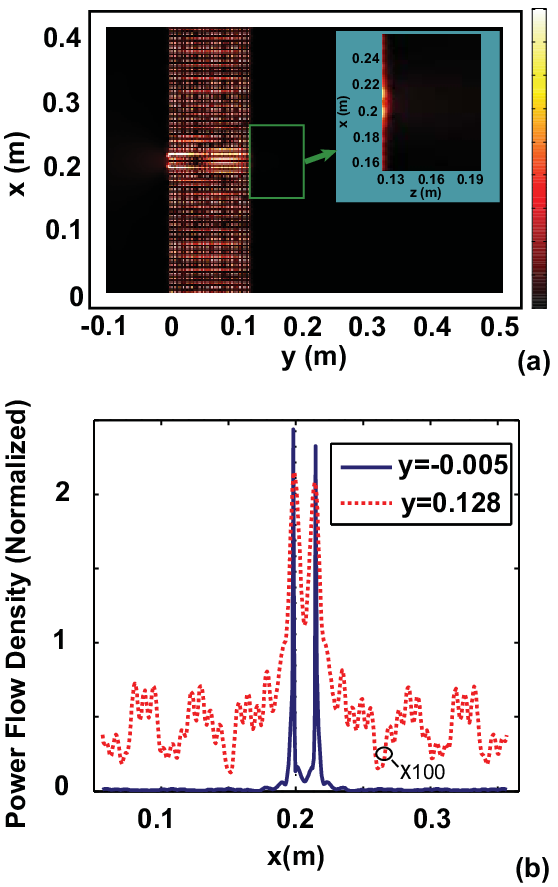, width=0.7\columnwidth,clip}
\end{center}
\caption{Sub-diffraction by a slab-shape hyperlens made of rod-type dielectric metamaterial. (a)PFDD; (b) Cross section draws of PFDD at $y$=-0.005m and $y$=0.128m. In the simulation, $f\approx$4.81GHz, $a$=1mm, $l$=7mm, $h$=6.4mm, while the interval of the two sources is 16mm. In (b), the data at $y$=0.128m is normalized by a factor of 100 to have a better comparison.}
\label{fig4}
\end{figure}

Again, BST cylinders with 1~mm radius are adopted to construct a metamaterial with $l$=7 mm and $h$=6.4 mm. In this way, a 0.128 m wide slab hyperlens is designed, see Fig.~\ref{fig4}(a). In the simulation, two point sources with a separation of 16 mm are placed in front of the hyperlens at $y$=-0.005~m. The simulated power flow density distribution around 4.81~GHz is shown in Fig.~\ref{fig4}(a), and indeed we observe two images at the interface. For a clearer visualization, panel~\ref{fig4}(b) shows cross-section cuts of the PFDD at $y$=-0.005~m and $y$=0.128~m. Keeping the radiation wavelength in mind, the ability to distinguish the two sources and the image indicates sub-diffraction imaging by the MMs hyperlens.

In conclusion, we have proposed a way to design anisotropic negative-index MMs by means of pure dielectric rod-type rectangular lattices. Results are supported by a homogenization approach as well as by numerical calculation and simulations. From our theoretical analysis, the effective dispersion is represented by closed form expressions like a realistic anisotropic media existing in nature.

With the anisotropic MMs proposed here, we have constructed a far field refocusing slablens and a subdiffraction hyperlens. Since the effective constitutive parameters play the key role in achieving the EM phenomena, anisotropic rod-type MMs could also be used to some other applications, e.g. constructing an invisibility cloak or transformation medium. Moreover, our study in this paper points out flexible design of dielectric rod-type MMs, which avoids the loss and scaling issue of metallic structures in realizing negative properties, indicating many potential applications in terahertz and optical range in future.

\emph{Acknowledgments.} This work was sponsored by NSFC (No.60531020), 863 Project (No.2009AA01Z227), NCET-07-0750 and the National Key Laboratory Foundation (Nos.9140C5304020901 and Nos.9140C5304020704). L.~P. and N.~A.~M. also acknowledge financial support by the Danish National Advanced Technology Foundation (grant no: 004-2007-1).

\newpage

%\bibitem{Kurs:2007}
%A. Kurs, A. Karalis, R. Moffatt, J.~D. Joannopoulos, P. Fisher, and M. Soljacic, Science {\bf 317}, 83 (2007).

\newpage

\end{document}